\documentclass{aastex63}

\usepackage{setspace}

\usepackage{subfigure}
\usepackage{comment}

\begin{document}

\newcommand{\average}[1]{\ensuremath{\langle#1\rangle} }

\title{ Effects of Hoyle state de-excitation on  $\nu p$--process nucleosynthesis and Galactic chemical evolution
}

\author{Hirokazu Sasaki}

\affiliation{Division of Science, National Astronomical Observatory of Japan, \\
2-21-1 Osawa, Mitaka, Tokyo 181-8588, Japan}

\author{Yuta Yamazaki}
\affiliation{Division of Science, National Astronomical Observatory of Japan, \\
2-21-1 Osawa, Mitaka, Tokyo 181-8588, Japan}
\affiliation{Graduate School of Science, The University of Tokyo, \\
7-3-1 Hongo, Bunkyo-ku, Tokyo 113-033, Japan}

\author{Toshitaka Kajino}
\affiliation{Division of Science, National Astronomical Observatory of Japan, \\
2-21-1 Osawa, Mitaka, Tokyo 181-8588, Japan}
\affiliation{Graduate School of Science, The University of Tokyo, \\
7-3-1 Hongo, Bunkyo-ku, Tokyo 113-033, Japan}
\affiliation{School of Physics, and International Research Center for Big-Bang Cosmology and Element Genesis, \\
Beihang University, Beijing 100183, China}
\affiliation{Peng Huanwu Collaborative Center for Research and Education, Beihang University, Beijing 100183, China}
\author{Grant J. Mathews}
\affiliation{Department of Physics and Astronomy, Center for Astrophysics, University of Notre Dame, Notre Dame, IN 46556, USA}











\setstretch{1.5}
\large
\begin{abstract}


The partcle-induced hadronic de-excitation of the Hoyle state in $^{12}$C induced by inelastic scattering in a hot and dense plasma can enhance the triple-alpha reaction  rate. This prevents the production of heavy nuclei within the neutrino-driven winds of core-collapse supernovae and raises a question  as to the contribution of proton-rich neutrino-driven winds as the origin of $p$--nuclei in the solar system  abundances. Here we study  $\nu p$-process nucleosynthesis in proton-rich neutrino-driven winds relevant to the production of $^{92,94}\mathrm{Mo}$ and $^{96,98}\mathrm{Ru}$ by considering such particle-induced de-excitation. We show that the enhancement of the triple-alpha reaction rate induced by neutron inelastic scattering hardly affects the $\nu p$-process,  while  the proton scattering contributes to the nucleosynthesis in proton-rich neutrino-driven winds at low temperature. The  associated enhanced triple-alpha reaction rate decreases the production of  $^{92,94}\mathrm{Mo}$ and $^{96,98}\mathrm{Ru}$ in a wind model of ordinary core-collapse supernovae. On the other hand, the abundances of these $p$--nuclei  increase in an energetic hypernova wind model.  Hence, we calculate the galactic chemical evolution of $^{92,94}\mathrm{Mo}$ and $^{96,98}\mathrm{Ru}$ by taking account of both contributions from core-collapse supernovae and hypernovae. We show that the hypernova $\nu p$-process can enhance the calculated solar isotopic fractions of $^{92,94}\mathrm{Mo}$ and $^{96,98}\mathrm{Ru}$  and make a significant impact on the GCE of $p$--nuclei regardless of the particle-induced Hoyle state de-excitation. 


\end{abstract}

\section{Introduction} 
\label{sec:introduction}
The excited $0^{+}$ state of $^{12}\mathrm{C}$ at 7.65 MeV  (the so-called ``Hoyle state") can resonantly enhance the reaction rate of the triple-alpha ($3\alpha$) process essential for nucleosynthesis inside stars \citep{Burbidge1957}. In a hot and dense plasma, inelastic scatterings of background particles can induce the hadronic de-excitation of the Hoyle state to the ground state or to the excited $2^{+}$ state at 4.44 MeV. This enhances the $3\alpha$ reaction rate \citep{Beard2017}. In this context, the enhancement was calculated based on the statistical Hauser-Feshbach model \citep{Beard2017}, and the contribution of neutrons was recently measured in a neutron inelastic scattering experiment \citep{Bishop2022}. 

The $3\alpha$ reaction is crucial for the nucleosynthesis of heavy elements inside explosive astrophysical sites such as core-collapse supernovae (CCSNe) and neutron star mergers. In particular, the $\nu p$--process nucleosynthesis \citep{Frohlich2006}, that  can occur  within the proton-rich neutrino-driven winds of core-collapse supernovae is sensitive to the uncertainty of the $3\alpha$ reaction rate \citep{Wanajo2011,Nishimura2019}. The $\nu p$--process is induced by the absorption of electron antineutrinos on free protons, $p(\bar{\nu}_{e},e^{+})n$ \citep{Frohlich2006,Pruet2006,Wanajo2006}. The free neutrons produced via the neutrino absorption allow for the production of heavier elements beyond the waiting point nucleus $^{64}$Ge and other bottleneck nuclei through $(n,p)$ reactions instead of  slower $\beta^{+}$ decays. The $\nu p$--process  is affected also by various nuclear reactions rates.  Among them  the  $^{59}\mathrm{Cu}(p,\alpha)^{56}\mathrm{Ni}$ and $^{7}\mathrm{Be}(\alpha,\gamma)^{11}\mathrm{C}$ were recently measured in nuclear experiments \citep{Randhawa2021,Psaltis2022PRL}.

The $\nu p$--process can produce large numbers of $p$-nuclei that cannot be synthesized through either the slow ($s$-) or rapid ($r$-) neutron capture processes.  Most $p$-nuclei can be produced in the $\gamma$-process \citep{Woosley1978,Hayakawa2004,Hayakawa2008} induced by successive photodisintegration reactions on heavier isotopes.  However,  calculations of the galactic chemical evolution (GCE) of  abundant $p$-isotopes such as $^{92,94}\mathrm{Mo}$ and $^{96,98}\mathrm{Ru}$ in models that only include  the $\gamma$-process in the outer layers of both thermonuclear supernovae (SNe Ia) and ordinary core-collapse supernovae (SNe II) drastically underestimate the solar isotopic fractions  \citep{Travaglio2018}. 

Moreover, the molybdenum isotopic anomalies in meteorites indicate that the $p$-isotopes $^{92,94}\mathrm{Mo}$ and the $r$-isotope $^{100}\mathrm{Mo}$ are synthesized in the same star but by different processes \citep{Dauphas2002,Budde2016,Poole2017}.  A $\nu p$--process in core-collapse supernovae where the $r$-process also occurs could meet such a requirement and be a candidate site for the production of  these abundant $p$-isotopes $^{92,94}\mathrm{Mo}$ and $^{96,98}\mathrm{Ru}$. In particular,  a  strong $\nu p$--process is possible in the proton-rich neutrino-driven winds of very energetic hypernovae (HNe)  \citep{Fujibayashi2015}. Such a HN $\nu p$--process significantly increases the elemental abundances of Mo and Ru at low metallicity $[\mathrm{Fe/H}]<-2$ \citep{Sasaki2022ApJ}. 

 Recently, \cite{Jin2020} reported that the enhanced $3\alpha$ reaction induced by the hadronic de-excitation  of  the Hoyle state should increase the seed nuclei for the production of heavy elements and suppress the $\nu p$--process.  This raises a question as to the impact of the $\nu p$-process on the solar abundances of $^{92,94}$Mo and $^{96,98}$Ru. This seems to be true for a neutrino-driven wind model in SNe II with small entropy per baryon and a large expansion timescale. However, the contribution  to the $\nu p$--process of such particle-induced Hoyle state de-excitation in HNe with a massive  proto-neutron star (PNS) and large neutrino luminosities is still uncertain \citep{Fujibayashi2015}. Neutrino-driven winds in such HNe have larger entropy and  a shorter expansion timescale than the SN II wind model. The $\nu p$--process in SNe II hardly affects the GCE due to the  relatively small production yield of $p$-nuclei while the HN $\nu p$--process can dominantly contribute \citep{Sasaki2022ApJ}. Observational quantities such as the solar isotopic fractions and elemental abundances can be affected by the Hoyle state effect in the HN $\nu p$--process. 

For the present work, we calculate the enahnced $3\alpha$ reaction in the $\nu p$--process by using both SN II and HN wind models. Then, we carry out the GCE calculation with the calculated nuclear yields  to
demonstrate how the particle-induced Hoyle state de-excitation in the HN $\nu p$--process affects the GCE of Mo and Ru.

\section{Methods}
\label{sec:method}

\subsection{Models of neutrino-driven winds in supernova and hypernovae}

We need hydrodynamic quantities and neutrino fluxes to calculate the $\nu p$--process nucleosynthesis  within  neutrino-driven winds. We calculate the temperature and baryon density profiles of the neutrino-driven wind  based upon a model for general-relativistic steady-state, spherically symmetric trajectories \citep{Otsuki2000}. The radius of  the PNS is taken to be $R_{\mathrm{PNS}}=15$ km.  The baryon density  near the PNS radius is taken to be $\rho_{0}=10^{11}\mathrm{g}/\mathrm{cm}^{3}$, and the temperature at the PNS radius is determined by  the condition $\dot{q}=0$, where  $\dot{q}$ is the net heating rate  from  neutrino interactions \citep{Otsuki2000}. We assume that the neutrino luminosity $L_{\nu}$ is independent of neutrino species,  and that these neutrinos obey Fermi-Dirac distributions on the surface of the PNS,  with the neutrino mean energies  fixed to $\average{E_{\nu_{e}}}=$ 13.1 MeV, $\average{E_{\bar{\nu}_{e}}}=$ 15.7 MeV, and $\average{E_{\nu_{X}}}=$ 16.3 MeV. With these neutrino parameters, we  have calculated the rates of neutrino-induced reactions and the electron fraction inside the neutrino-driven winds. The calculated electron fraction is used to determine mass fractions of initial neutrons and protons for the  subsequent nuclear network calculation. 

To study the effect of the particle-induced Hoyle state de-excitation, we prepare wind trajectories of both ordinary SN II and energetic HN models with different values of the PNS mass $M_{\mathrm{PNS}}$ and $L_{\nu}$ as shown in Table \ref{tab:wind models}. These parameters for the SN II model are typical values for the late stages of the explosion (e.g. see \citet{Burrows2020,Nagakura2021}). We assume that a collapse of a rapidly rotating massive star is associated with the energetic explosion mechanism of HNe. For the HN wind model, we consider a proton-rich neutrino-driven wind blown off from the massive PNS toward the polar region before the black hole formation as in \citet{Fujibayashi2015} by employing a large PNS mass $(M_{\mathrm{PNS}}=3M_{\odot})$ and a large neutrino luminosity $(L_{\nu}=10^{53} \mathrm{erg/s})$ as seen in neutrino radiation hydrodynamic simulations \citep{Sekiguchi2012,Fujibayashi2021}.

\subsection{Enhanced $3\alpha$ reaction rate due to Hoyle state de-excitation}

We execute the nuclear network calculation following the numerical  setup of \cite{Sasaki2017,Sasaki2022ApJ}. We use  the LIBNUCNET reaction network engine \citep{Meyer2007} with nuclear reaction rates  from the  JINA Reaclib database \citep{Cyburt2010}. We  have included reaction rates for neutrino absorption and $e^{\pm}$ capture. However, we ignore the contribution from  neutrino oscillations \citep{ko2020,ko2022}. The enhanced $3\alpha$ reaction rate owing to the induced Hoyle state de-excitation is given by \cite{Jin2020},
\begin{eqnarray}    \lambda_{3\alpha}&=&\lambda_{3\alpha}^{(0)}\left\{
    1+Y_{n}\rho_{6}f_{n}(T_{9})+Y_{p}\rho_{6}f_{p}(T_{9})
    \right\},\label{eq:enhanced triple-alpha}\\
f_{n}(T_{9})&=&75.1e^{-T_{9}}+88.7,\label{eq:fn}\\
f_{p}(T_{9})&=&0.03680-1.667T_{9}+2.350T_{9}^{2}-0.2911T_{9}^{3}+0.01160T_{9}^{4},
\end{eqnarray}
where $\lambda_{3\alpha}^{(0)}$ is the $3\alpha$ reaction rate without the enhancement \citep{Fynbo2005}, $\rho_{6}$ is the baryon density in  units  of $10^{6}  ~\mathrm{g~cm^{-3}}$, and $T_{9}$ is the temperature in  units of $10^{9}$ K.  The quantities $Y_{n}$ and $Y_{p}$ are the number abundance fractions of free neutrons and protons, respectively. The second and third terms on the right-hand side of Eq.~(\ref{eq:enhanced triple-alpha}) are the contributions from neutron and proton inelastic  scattering, and the values of $f_{n}$ and $f_{p}$ are determined by fitting to the statistical Hauser-Feshbach calculation of \citet{Beard2017}.  This enhanced $3\alpha$ reaction is utilized for the network calculations  in the present SN and HN nucleosynthesis models.

\subsection{Model for Galactic chemical evolution}

 We adopt the GCE model of \citet{timmes1995} which reproduces reasonably well the chemical evolution of the light elements from hydrogen to zinc as well as the model of  \citet{ko2020}. The adopted model \citep{timmes1995} has already been successfully applied to the GCE of the intermediate-to-heavy mass nuclei including the $r$--process contributions from magneto-hydrodynamic jet SNe, collapsars and binary neutron star mergers as well as the neutrino-driven wind in core collapse SNe \citep{yamazaki2021,yamazaki2022} and the $\nu p$--process contributions from Type II SNe and Hypernovae \citep{Sasaki2022ApJ}. The latter study \citep{Sasaki2022ApJ} includes not only the $\nu p$--process, but the $\gamma$-- ($p$--), $s$--, and $r$--processes in addition to the $\nu p$--process.
In the present calculations, we follow the same numerical setup except for the input data of the HN $\nu p$--process as discussed in the previous section. 

We demonstrate the impact of the $\nu p$--process on the GCE calculation with the enhanced $3\alpha$ reaction rate in Eq.~(\ref{eq:enhanced triple-alpha}). We focus on the GCE of $^{92,94}\mathrm{Mo}$ and $^{96,98}\mathrm{Ru}$ whose total solar isotopic fractions are as high as $24.1\%$ and $7.4\%$, respectively.  A GCE calculation that only includes the $\gamma$--process underestimates such large solar abundances and the $\nu p$--process potentially resolves this underestimation problem. The HN $\nu p$--process can be a main contributor to the GCE of $^{92,94}\mathrm{Mo}$ and $^{96,98}\mathrm{Ru}$ \citep{Sasaki2022ApJ} so that the effect of the particle-induced Hoyle state de-excitation on   GCE could be  well demonstrated by considering the HN $\nu p$--process. 

 We consider the $\gamma$--process in both SNe Ia and II and the $\nu p$--process in both SNe II and HNe as the astrophysical sources of $p$-nuclei. We use the HN wind model in Table~\ref{tab:wind models} as the fiducial proton-rich neutrino-driven winds in HNe with Eq.~(\ref{eq:enhanced triple-alpha}). In the HN $\nu p$--process, the yield of the nucleus $i$ is estimated by $X_{i}\dot{M}\tau_{NS}$ where $X_{i}$ and $\dot{M}$ are  a mass fraction of $i$ and the mass ejection rate inside the HN wind, and $\tau_{NS}=1$ s is a typical lifetime of the massive PNS \citep{Fujibayashi2015}. The progenitor mass for the HN model is set to $100$ M$_{\odot}$ as in \citet{Sasaki2022ApJ}. 

\vspace{0.5cm}
\section{Results and discussions}
\label{sec:results}

\subsection{Hydrodynamic and neutrino properties}

Table~\ref{tab:wind models} shows the hydrodynamic quantities characterizing properties of the wind models such as the expansion timescale $\tau_{\mathrm{dyn}}$, the entropy per baryon $S$, the initial electron fraction for the network calculation $Y_{e}^{(0)}$, and the mass ejection rate $\dot{M}$. The values of $\tau_{\mathrm{dyn}}$ and $S$ are calculated at a high temperature 
before the production of heavy elements as in \citep{Sasaki2022ApJ}. The entropy for the HN wind model is higher than that of the SN II wind model due to the massive PNS mass. The small expansion timescale in the HN model  originates from the large neutrino luminosity. Such properties of neutrino-driven winds are consistent with the results of \citet{Otsuki2000,Fujibayashi2015}. $Y_{e}^{(0)}$ corresponds to the electron fraction at the beginning of the nuclear network calculation ($T_{9}=10$). The value of $Y_{e}^{(0)}$ is larger than 0.5 for the proton-rich neutrino-driven winds, and almost the same in both wind models because we use the same neutrino energies for the neutrino distributions. $Y_{e}^{(0)}$ becomes larger when the difference between the mean neutrino energies, $\average{E_{\bar{\nu}_{e}}}-\average{E_{\nu_{e}}}$ is small. $\dot{M}$ is determined from a supersonic wind solution, and the value increases with the neutrino luminosity. In an ordinary SN explosion, the entropy of the wind increases with the decrease of neutrino luminosity in the later explosion phase ($t>1$ s). Although a high entropy is favorable for the production of heavy elements, their total   yields in the later wind trajectories are not so large due to the small $\dot{M}$ \citep{Sasaki2022ApJ}. The massive PNS and the large neutrino luminosity in the HN model simultaneously enable both a high entropy and a  large mass ejection rate.

\begin{table}
\begin{center}
\small
\begin{tabular}{|c c c|c c c c|} \hline
Model&$M_{\mathrm{PNS}}$($M_{\odot}$)&$L_{\nu}$($10^{51}$erg/s)&$\tau_{\mathrm{dyn}}$(ms)&$S$($k_{B}\mathrm{nuc}^{-1}$)&$Y^{(0)}_{e}$&$\dot{M}(M_{\odot}/\mathrm{s})$
\\ \hline
SN II&$1.4$&10&16.9&55.2&0.534&5.62$\times10^{-4}$\\ 
HN&$3$&100&5.18&125&0.531&4.46$\times10^{-3}$\\ \hline
\end{tabular}
\end{center}
\small
\caption{Properties of SN II and HN wind models.}\label{tab:wind models}
\end{table}

\subsection{$\nu p$--process nucleosynthesis}

\begin{figure}
\centering
\subfigure{%
    \includegraphics[clip, width=0.5\linewidth]{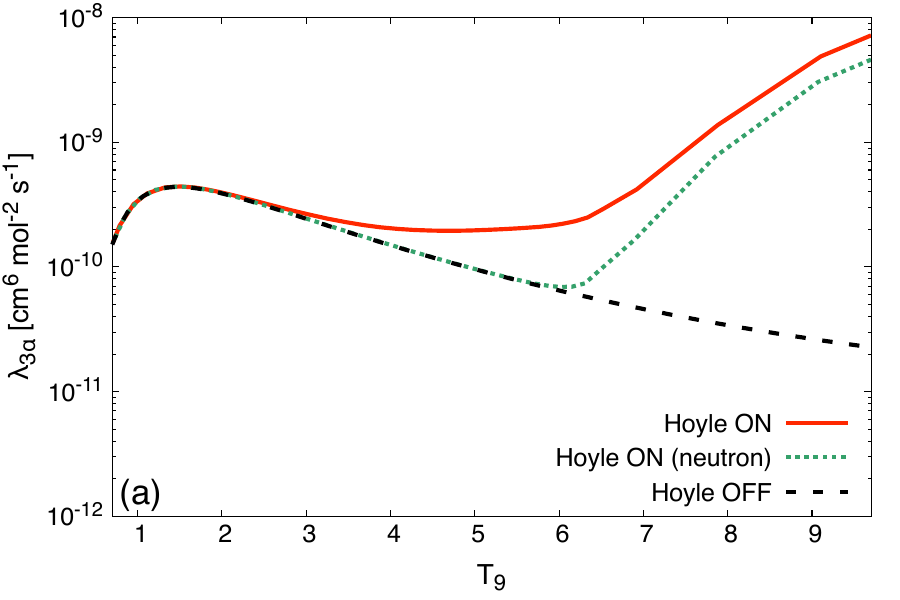}
    }
\subfigure{%
    \includegraphics[clip, width=0.5\linewidth]{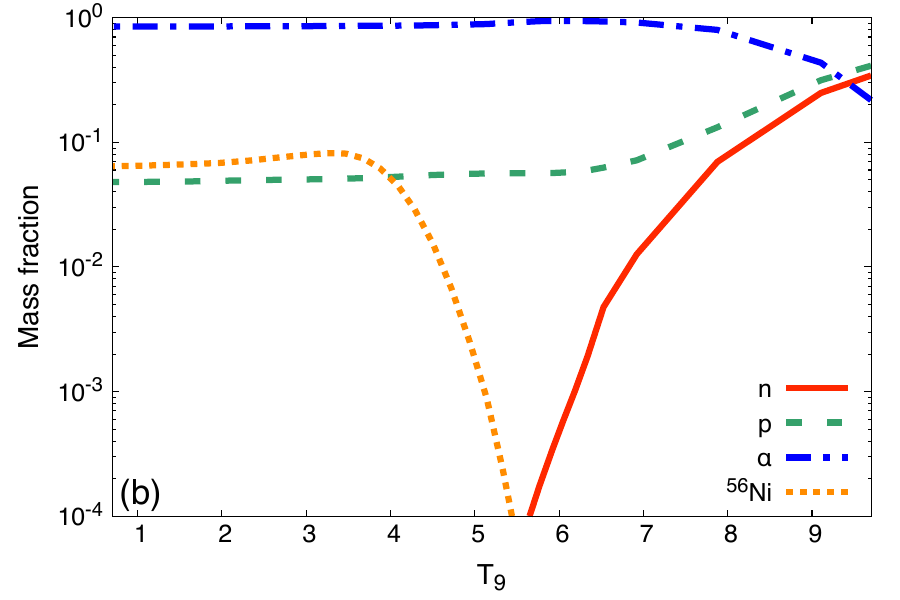}
    }\caption{(a) The calculated $3\alpha$ reaction rates in the SN II wind model. The solid and dashed  lines show the results of Eq.~(\ref{eq:enhanced triple-alpha}) with ($f_{n},f_{p}\neq0$) and without ($f_{n}=f_{p}=0$) the effects of particle-induced Hoyle state de-excitation. The dotted line shows the contribution from only neutron scattering in Eq.~(\ref{eq:enhanced triple-alpha}) ($f_{n}\neq0,f_{p}=0$). (b) The mass fractions of neutrons, protons, $\alpha$-particles, and $^{56}\mathrm{Ni}$ within the SN II wind model  including the effects of Hoyle state de-excitation (Hoyle ON) in Fig.~\ref{fig:reaction rates SN}(a).}
    \label{fig:reaction rates SN}
\end{figure}


\begin{figure}
\centering
\subfigure{%
    \includegraphics[clip, width=0.5\linewidth]{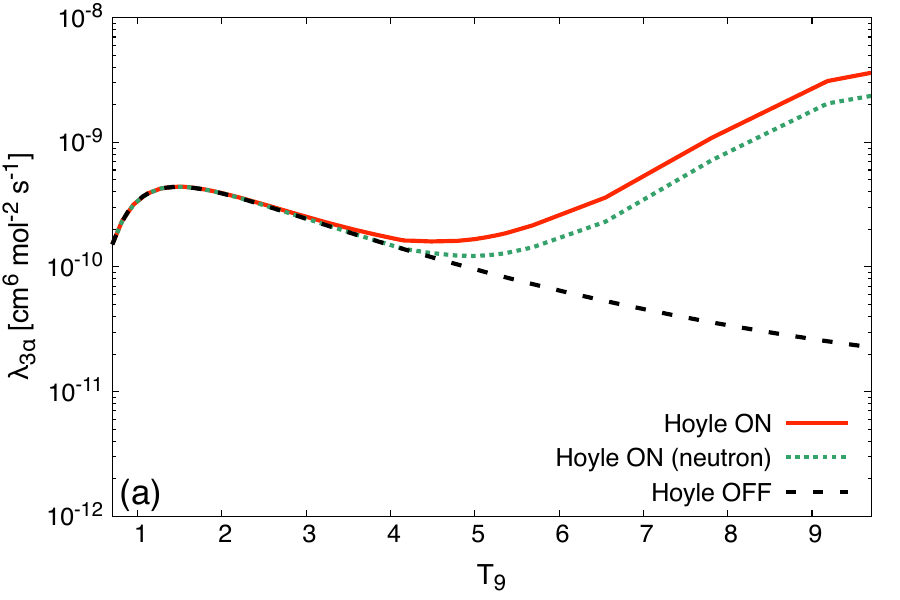}
    }
\subfigure{%
    \includegraphics[clip, width=0.5\linewidth]{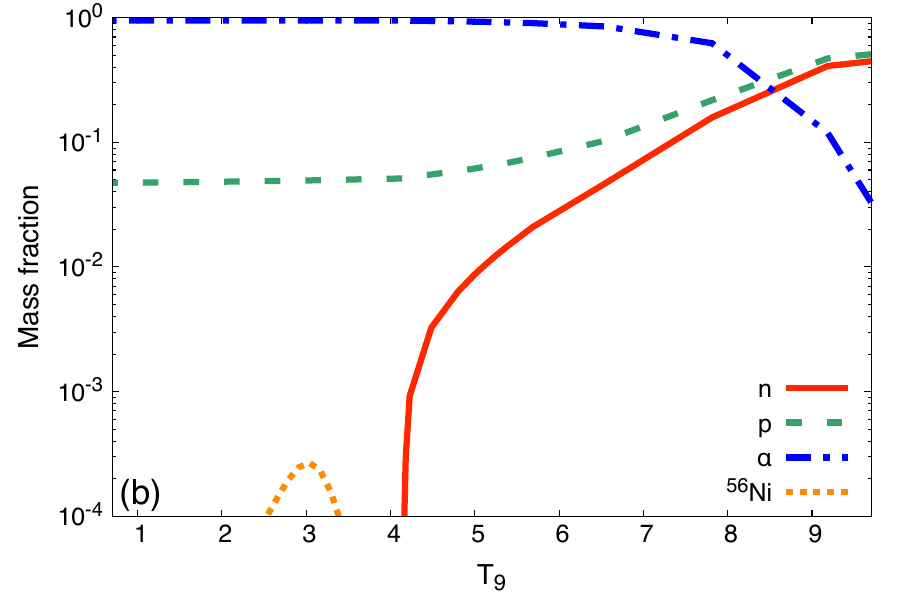}
    }\caption{The results of the HN wind model as in Fig.~\ref{fig:reaction rates SN}.}\label{fig:reaction rates HN}
\end{figure}

Figure~\ref{fig:reaction rates SN}(a) shows  the $3\alpha$ reaction rates used for the $\nu p$--process calculations in the SN II wind model.  The solid and dashed lines are respectively the rates with and without the particle-induced Hoyle state de-excitation of neutron and proton scatterings. To see the contribution of the neutron-induced enhancement,
we  obtained the dotted line by setting $f_{p}=0$ in Eq.~(\ref{eq:enhanced triple-alpha}). The $3\alpha$ reaction rate is enhanced by more than  a factor of $100$ at higher temperatures ($T_{9}>9$) due to the large baryon density $\rho_{6}$ and the  large abundance of free nucleons  (Fig.~\ref{fig:reaction rates SN}(b)). The dashed and dotted lines are almost identical  at $T_{9}<6$ and the contribution from the neutron scattering becomes negligible.  This is because the mass fraction of free neutrons significantly decreases with the production of seed nuclei around $^{56}\mathrm{Ni}$ through the $\alpha$-capture reactions as shown in Fig.~\ref{fig:reaction rates SN}(b). Then, the enhancement of the $3\alpha$ reaction is mainly caused by  proton scattering in the temperature range $2<T_{9}<6$ due to the freeze out of the protons as in Fig.~\ref{fig:reaction rates SN}(b). Finally, as the baryon density decreases, the enhancement is negligible in the low-temperature region, $T_{9}<2$. 

Figure~\ref{fig:reaction rates HN} shows the enhanced $3\alpha$ reaction rates and the evolution of mass fractions in the HN wind model. The results are similar to the case of  the SN II wind model. The proton scattering only contributes to the enhancement in Eq.~(\ref{eq:enhanced triple-alpha}) for $T_{9}<4$. This is due to the freeze out of free protons and decreasing free neutrons. The large entropy per baryon of the HN wind model results in  a small amount of seed nuclei such as $^{56}$Ni and a large production of heavy elements through the $\nu p$--process.

\begin{figure}
\subfigure{%
    \includegraphics[clip, width=0.5\columnwidth]{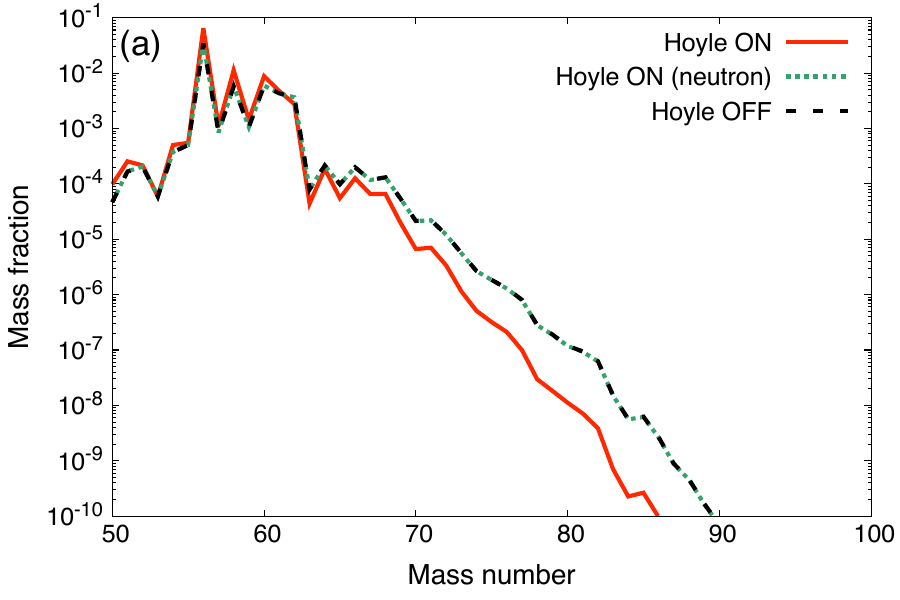}}%
\subfigure{%
    \includegraphics[clip, width=0.5\columnwidth]{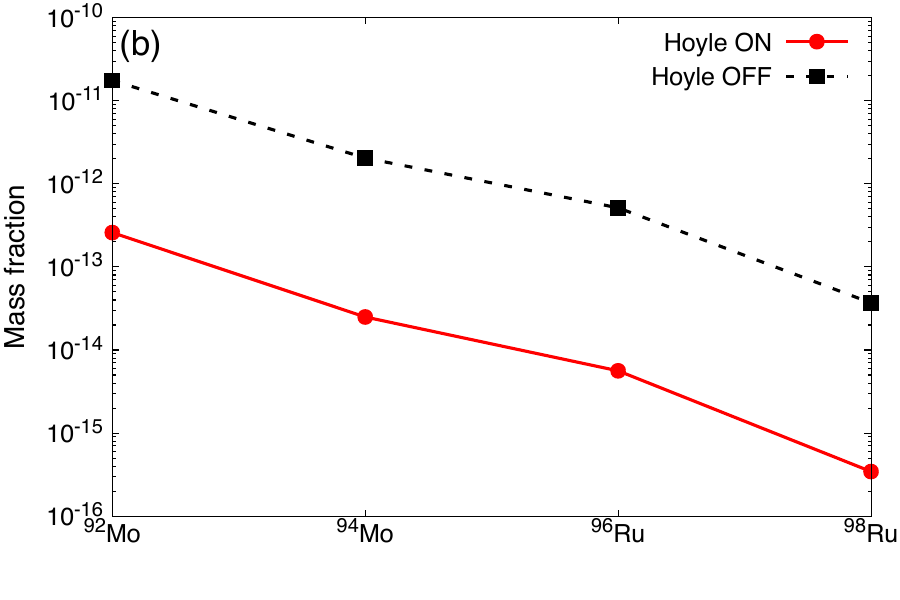}}%
    \caption{Effect of the particle-induced Hoyle state de-excitation on the calculated mass fractions for (a) nuclei with $A>50$ and (b) $^{92,94}\mathrm{Mo}$ and $^{96,98}\mathrm{Ru}$ in the SN II wind model. The dotted line in (a) shows the result assuming $f_{p}=0$ in Eq.(\ref{eq:enhanced triple-alpha}).}\label{fig:mass fractions in SN}
\end{figure}

\begin{figure}
\subfigure{%
    \includegraphics[clip, width=0.5\columnwidth]{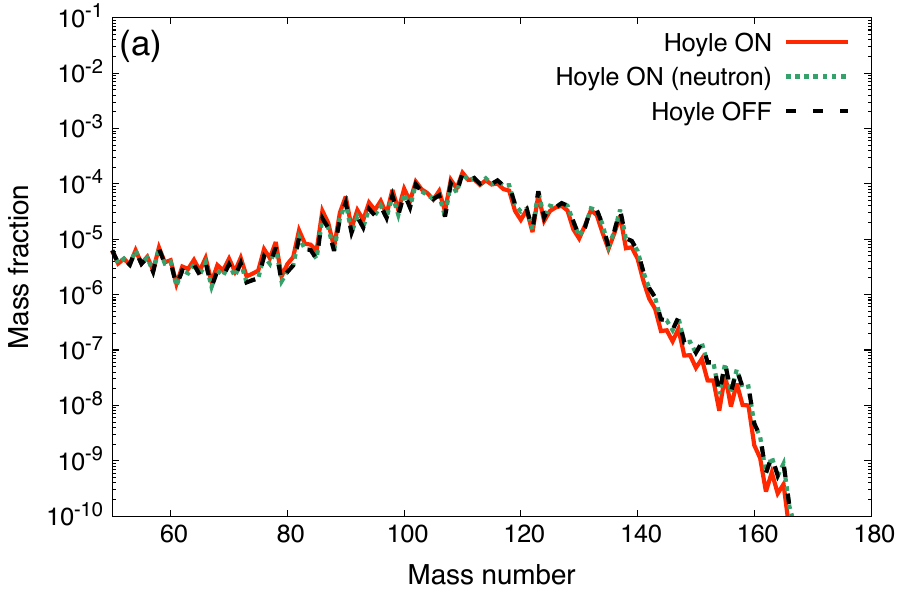}}%
\subfigure{%
    \includegraphics[clip, width=0.5\columnwidth]{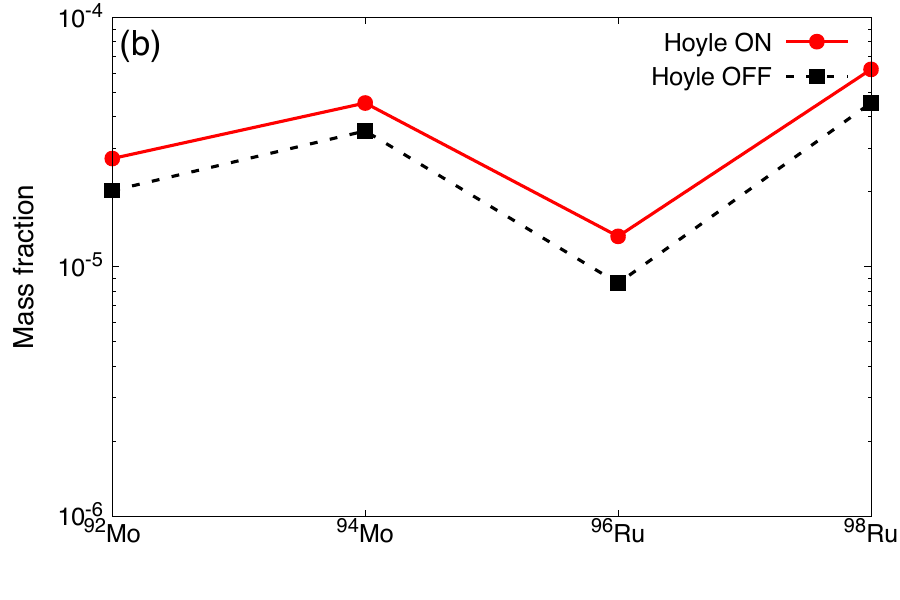}}%
    \caption{The calculated mass fractions in the HN wind model as in Fig.~\ref{fig:mass fractions in SN}.}\label{fig:mass fractions in HN}
\end{figure}

Figure~\ref{fig:mass fractions in SN}(a) shows the effect of the particle-induced Hoyle state de-excitation on the nuclear mass fractions  of final abundances for various nuclei in the SN II wind model.  The solid and dashed lines show the results with  and without the Hoyle state effect, respectively, and the difference between them is prominent  for nuclei  with  $A>60$. This is because the enhanced $3\alpha$ reaction suppresses the production of heavy elements through the $\nu p$--process. Figure~\ref{fig:mass fractions in SN}(b) shows the mass fractions of $^{92,94}\mathrm{Mo}$ and $^{96,98}\mathrm{Ru}$ in the SN II wind model with and without the enhancement of Eq.~(\ref{eq:enhanced triple-alpha}). The results with  the enhanced $3\alpha$ reaction rate (solid line) are smaller than those without it (dashed line) by about  a factor of $100$.   Such significant suppression of the $\nu p$--process in the SN II wind model having $S\sim 50\ k_{\mathrm{B}}\mathrm{nuc}^{-1}, \tau_{\mathrm{dyn}}\sim 10\ \mathrm{ms}$ is consistent with the results of \citet{Jin2020}. 

The dotted line on Fig.~\ref{fig:mass fractions in SN}(a) shows the calculated mass fraction  obtained with the enhanced $3\alpha$  reaction rate ignoring the third term  on the right-hand side of Eq.~(\ref{eq:enhanced triple-alpha}) which is the contribution from proton-induced de-excitation. The dotted and dashed lines almost completely  overlap.  This indicates the negligible impact of  neutron scattering on the $\nu p$--process. As shown  by the dotted line on Fig.~\ref{fig:reaction rates SN},  neutron scattering induces an enhancement of the $3\alpha$ reaction rate at $T_{9}>6$. However, the $3\alpha$ reaction hardly affects  the  nuclear abundances in such a high-temperature region.  The enhancement of the $3\alpha$ reaction only contributes to the nuclear network calculation at $T_{9}<6$ where the free neutrons are consumed by the synthesis of seed nuclei (e.g. $^{56}$Ni) as in Fig.~\ref{fig:reaction rates SN}(b). We  note that the contribution from  neutron scattering should be negligible even if we use  the experimentally determined value of $f_{n}$ from \citet{Bishop2022}, which  turns out to be much smaller than that of Eq.~(\ref{eq:fn}).

The calculated mass fractions in the HN wind model are shown in Fig.~\ref{fig:mass fractions in HN}. The contribution  from  neutron scattering is negligible as in the case of the SN II wind  model. Hence,  there is no difference between the dashed and dotted lines on Fig.~\ref{fig:mass fractions in HN}(a). The $\nu p$--process in the HN wind model proceeds up to heavier elements than that of the SN II wind model because of the shorter $\tau_{\mathrm{dyn}}$ and higher $S$. 

The enhanced $3\alpha$ reaction rate decreases the production of heavy elements in the higher mass region ($A>140$) by 10--60 $\%$.  Also, the suppression is less significant than that of the SN II wind model due to the high entropy of the HN wind model. The value of  $\rho_{6}$ in Eq.~(\ref{eq:enhanced triple-alpha}) becomes small at a nearly fixed temperature of the $\alpha$-particle recombination $T_{9}\sim4$ in  the high entropy wind. Thus, the enhancement is small for the high entropy case \citep{Jin2020}. The enhanced $3\alpha$ reaction  rate decreases the ratio of free neutrons to the seed nuclei $\Delta_{n}$ from the $p(\bar{\nu}_{e},e^{+})n$ reaction \citep{Nishimura2019}. Such a decrease of $\Delta_{n}$ shifts the endpoint of the $\nu p$--process to lower masses and increases the mass fractions in  the  $A=60-110$ range. In particular, the mass fractions of $^{92,94}\mathrm{Mo}$ and $^{96,98}\mathrm{Ru}$  increase by 20--30$\%$ as shown in Fig.~\ref{fig:mass fractions in HN}(b). 

The enhanced $3\alpha$ reaction rate suppresses the production of heavy elements around the $\nu p$--process endpoint in both HN and SN II wind models. However, the HN wind model  produces sufficient heavy elements beyond $A>100$, while the SN II model does  not. Therefore, the role of the particle-induced Hoyle state de-excitation   in $^{92,94}\mathrm{Mo}$ and $^{96,98}\mathrm{Ru}$ depends  upon the wind models. The mass fractions of $^{92,94}\mathrm{Mo}$ and $^{96,98}\mathrm{Ru}$ in Fig.~\ref{fig:mass fractions in SN}(b) are too small to affect the GCE even without the suppression of the $\nu p$--process.  Hence, we will focus on neutrino-driven winds in HNe producing a large number of heavy elements, as shown in Fig.~\ref{fig:mass fractions in HN}(b), and we will demonstrate the effects of Hoyle state de-excitation on the solar abundances and the GCE of $^{92,94}\mathrm{Mo}$ and $^{96,98}\mathrm{Ru}$ in the next sections.

\subsection{ Comparison to Solar abundances}

\begin{figure}
\centering
\subfigure{%
    \includegraphics[clip, width=0.8\linewidth]{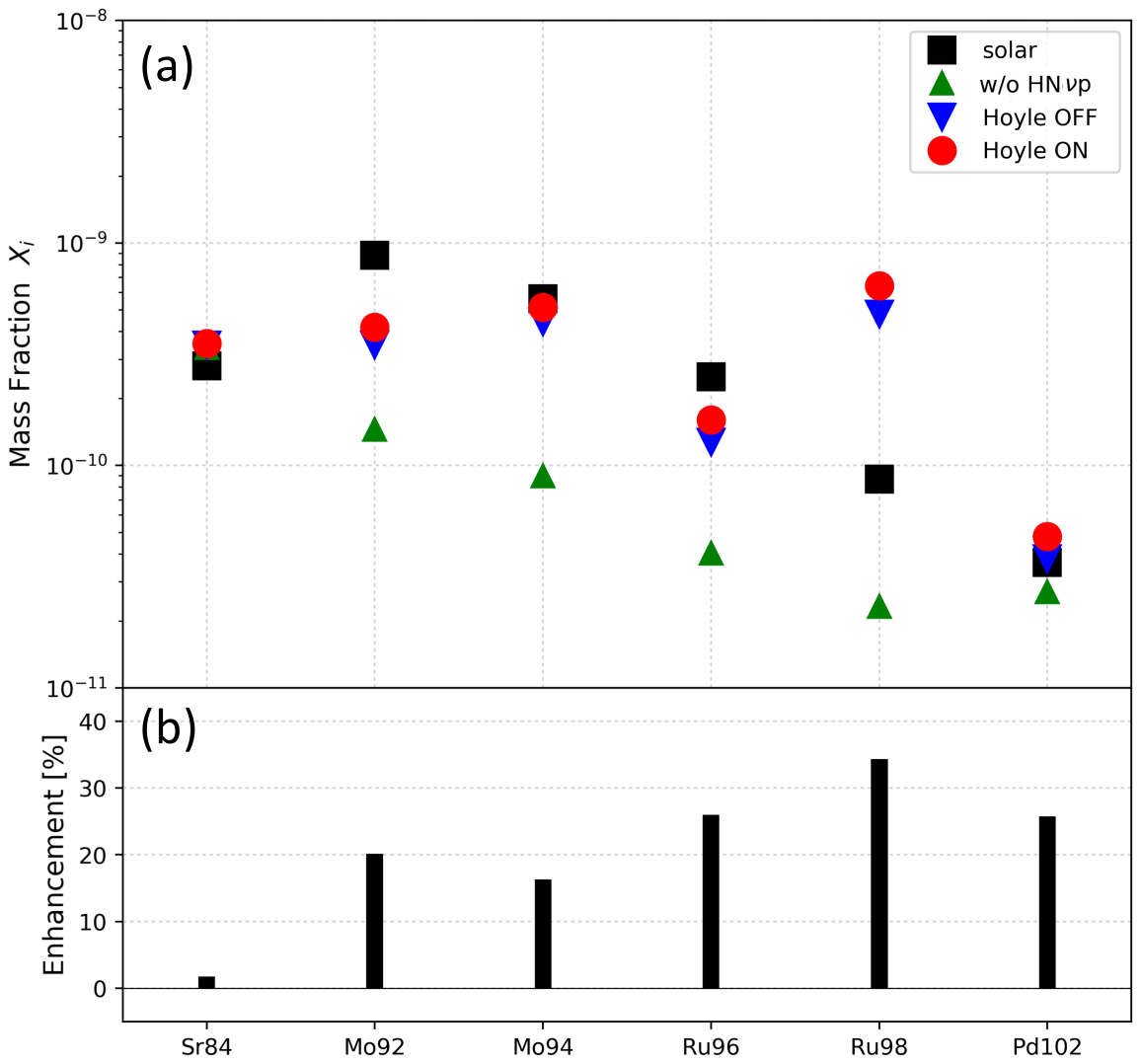}
    }\caption{(a)  Comparison between the calculated and observed solar abundances of $p$-nuclei at $[\mathrm{Fe}/\mathrm{H}]=0$. The square points are observational data \citep{Lodders_2003}. The upward-pointing  triangles show the calculated result without the HN $\nu p$--process.  The circles and downward-pointing triangles show respectively the results of the HN $\nu p$--process with and without the effects of particle-induced Hoyle state de-excitation. (b) The ratio of Hoyle OFF/Hoyle ON in the top panel.
    }\label{fig:solar abundance}
\end{figure}

\begin{figure}
\centering
\subfigure{%
    \includegraphics[clip, width=0.6\linewidth]{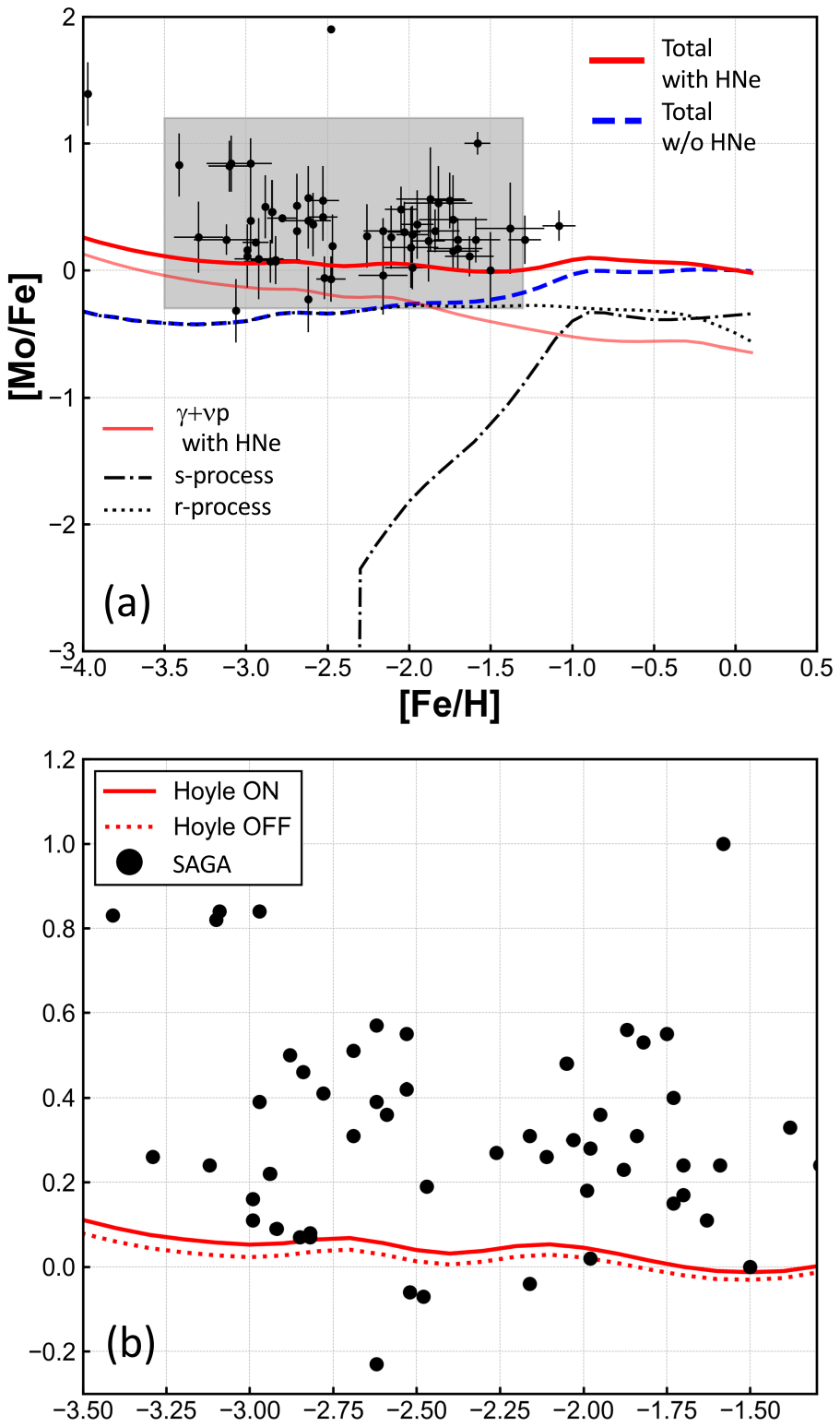}
    }\caption{(a) The evolution of the Mo elemental abundance. The circle points are observational data from the  SAGA database \citep{sagadatabase}. The thick solid line includes the HN $\nu p$--process with the  effects of particle-induced Hoyle state  de-excitation, while the thick dashed line  does not include the HN $\nu p$--process. Thin lines show the partial contributions from each process. (b) The impact of the Hoyle state effect on the total elemental abundance with the HN $\nu p$--process (thick solid line) is shown for the shaded region on the top panel  at $-3.50 < [\mathrm{Fe}/\mathrm{H}] < -1.30$.
    }\label{fig:mo abundance}
\end{figure}

\begin{figure}
\centering
\subfigure{%
    \includegraphics[clip, width=0.6\linewidth]{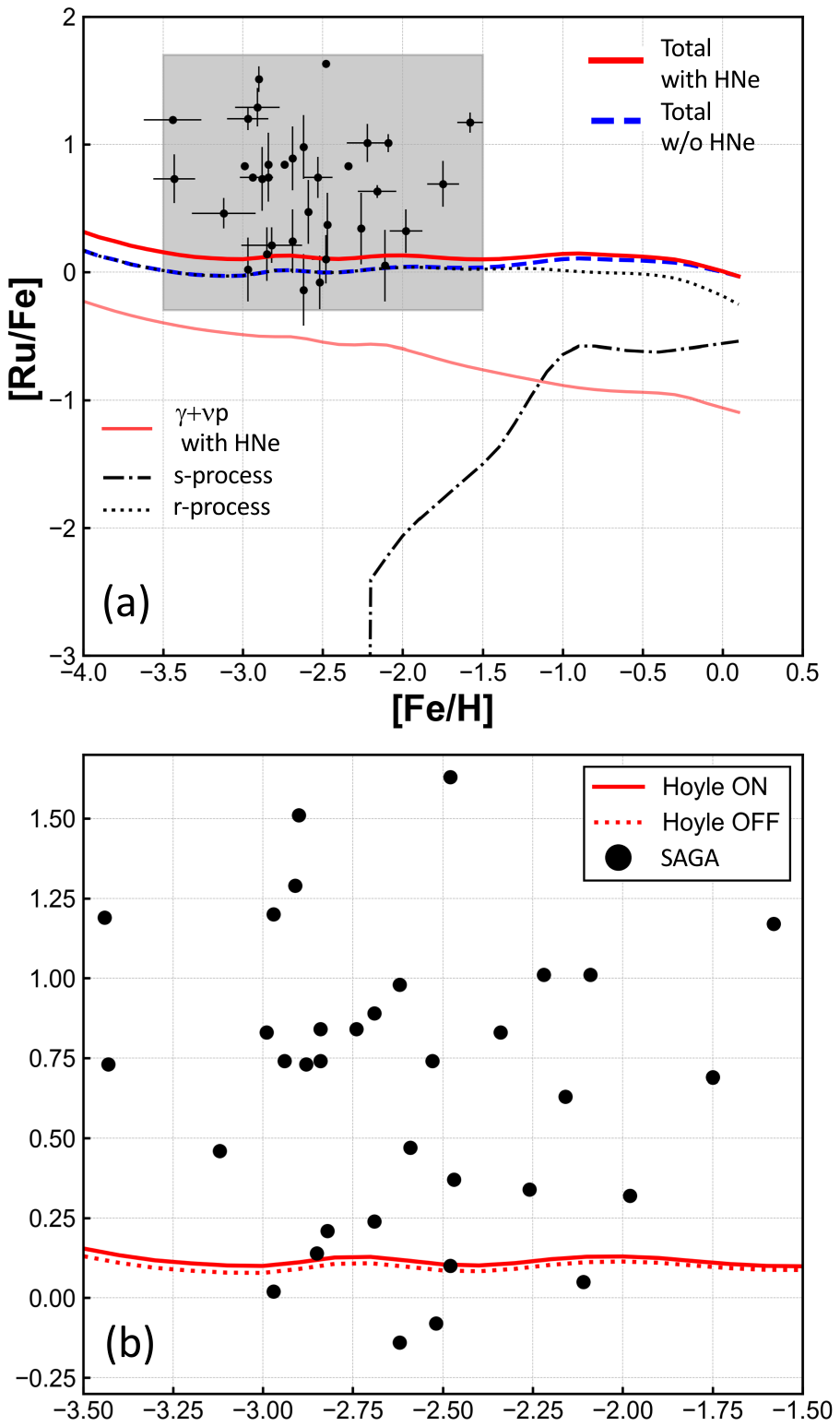}
    }\caption{The evolution of the Ru elemental abundance and the impact of the Hoyle state effect as in Fig.~\ref{fig:mo abundance}.}\label{fig:ru abundance}
\end{figure}

Figure.~\ref{fig:solar abundance}(a) shows the calculated solar abundances of $p$-nuclei in the $A=84$--$102$ range. These are  compared with observational data \citep{Lodders_2003}. All calculated results include the  contributions from the $\nu p$--process in SNe II and the $\gamma$--process in both SNe Ia and II. The result without taking into account of the HN $\nu p$--process (upward-pointing triangles) underestimates the solar abundances of $p$--nuclei  (square points). However, as shown  by the circles and  downward-pointing triangles, the HN $\nu p$--process can significantly contribute to the GCE of $p$--nuclei and increase the calculated solar abundances of $^{92,94}$Mo and $^{96,98}$Ru. The circles (Hoyle ON) and down-pointing triangles (Hoyle OFF) in Fig.~\ref{fig:solar abundance}(a) show the results including the HN $\nu p$--process with and without the effects of particle-induced Hoyle state de-excitation, respectively. For Hoyle ON and Hoyle OFF, we use the results of the solid and dashed lines in Fig.~\ref{fig:mass fractions in HN}, respectively. The HNe $\nu p$--process can increase the solar abundances of $^{92,94}$Mo and $^{96,98}$Ru irrespective of the Hoyle state effect.  This is consistent with the result of \cite{Sasaki2022ApJ} although \cite{Jin2020} suggests a  small contribution from the $\nu p$-process. The ratios of Hoyle OFF/Hoyle ON are shown in Fig.~\ref{fig:solar abundance}(b). Here, the abundances of $p$--nuclei are enhanced by up to about $30\%$. Although the suppression of $^{92,94}$Mo and $^{96,98}$Ru is due to the enhanced $3\alpha$ reaction rate as reported by \cite{Jin2020},   this occurs only in typical neutrino-driven winds of SNe II. On the other hand, such suppression is not necessarily found in the HN neutrino-driven wind based on the energetic supernova explosion model with a massive PNS and large neutrino luminosity as  given in \citep{Fujibayashi2015}. 

 The Hoyle ON points (circles) on Fig.~\ref{fig:solar abundance}(a) show the overestimation of $^{98}$Ru due to the large mass fraction of $^{98}$Ru in Fig.~\ref{fig:mass fractions in HN}. However, our calculation uses only one HN wind trajectory to estimate the abundances for the HN $\nu p$-process ignoring the time dependence of the neutrino-driven wind and simply multiplying by $\tau_{NS}=1$ s to obtain total integrated yields.  Hence, the overestimation might be resolved by integrating various neutrino-driven winds with  smaller PNS masses ($M_{\mathrm{PNS}}<3M_{\odot}$) over  time  until the black hole forms.

\subsection{Galactic chemical evolution of Mo and Ru isotopes}

Figure~\ref{fig:mo abundance} shows the calculated elemental abundances of Mo normalized  to the solar system values at $[\mathrm{Fe}/\mathrm{H}]=0$ and the observational data taken from the  SAGA database \citep{sagadatabase}. The thick solid line in Fig.~\ref{fig:mo abundance}(a) shows the result of Hoyle ON and the thick  dashed line shows the result without the contribution from the HN $\nu p$--process. These two lines indicate the significant enhancement of the elemental abundances. Moreover,  the HN $\nu p$--process improves the agreement between the GCE calculation with the observational data. Such impact of the HN $\nu p$--process was also found in \citep{Sasaki2022ApJ}. 

Figure~\ref{fig:mo abundance}(b) shows the impact of the particle-induced Hoyle state de-excitation on the elemental abundance in the shaded region of the top panel. The Hoyle state effect slightly increases the total elemental abundance of Mo with the HN $\nu p$--process. The thin solid, dash-dotted, and dotted lines on  Fig.~\ref{fig:mo abundance}(a) are partial contributions of the $(\gamma+\nu p)$--processes, the $s$--process, and the $r$--process, respectively. At low metallicity $[\mathrm{Fe}/\mathrm{H}]<-2$, the $\gamma$--process is negligible and the HN $\nu p$-process dominantly contributes to the total elemental abundance irrespective of the Hoyle state effect. 

The results for  the Ru elemental abundances are shown in Fig~\ref{fig:ru abundance}. The particle-induced Hoyle state de-excitation is almost unchanged.  The Ru elemental abundance and the contribution from the HN $\nu p$--process is less prominent than for Mo.  This is because the solar isotopic fractions of $^{96,98}$Ru are small ($7.4\%$) compared with those of $^{92,94}$Mo ($24.1\%$).  Also, the $r$--process is the main contributor  \citep{Bisterzo2014} to the Ru elemental abundance. To increase the calculated Ru elemental abundance at low metallicity,  an increased contribution from the $r$--process \citep{yamazaki2022} may be needed rather than the $\nu p$--process. This is because the overestimation of $^{98}$Ru in Fig.~\ref{fig:solar abundance}(a) would not be improved if  the yield of the $\nu p$--process was increased.


 Finally, we note several theoretical uncertainties in our calculation. First, 
we estimated the yield of the HN $\nu p$--process with  only one set of neutrino-driven wind trajectories.  However, the trajectories should also involve different explosion timescales and different progenitors leading to different massive PNSs. Also, nucleosynthesis calculations employing matter profiles obtained in neutrino radiation hydrodynamic simulations (e.g. \cite{Sekiguchi2012,Fujibayashi2021}) may provide more reliable yields of the $p$--nuclei. Another point is that, for simplicity we have ignored the contribution of neutrino oscillations  \citep{ko2020,ko2022} to the $\nu p$--process. In particular, fast flavor conversions \citep{fujimoto-nagakura2023} which have been actively studied in recent years can enhance the yields of $p$--nuclei \citep{Xiong2020}. Additionally, we have ignored the enhancement of the $3\alpha$ reaction induced by $\alpha$-particle inelastic scattering in the reaction network. Such reactions may have a non-negligible impact after the $\alpha$-rich freeze-out at low temperature as in Figs.~\ref{fig:reaction rates SN}(b) and \ref{fig:reaction rates HN}(b). Also, an $R$--matrix description \citep{R-matrix} incorporating experimental results could allow for a more sophisticated evaluation of the particle-induced Hoyle state de-excitation near the energy threshold.  This could provide a more accurate $3\alpha$ reaction rate for nucleosynthesis calculations. We  have employed theoretical estimates of nuclear masses for unstable nuclei. Nuclear masses obtained in recent mass measurements (e.g. \cite{Zhounaturephys2023})  could make the $\nu p$--process calculations more realistic and revise  the nuclear yields beyond the waiting-point nucleus $^{64}$Ge.

\vspace{0.5cm}

\section{Conclusion}
\label{sec:Conclusion}
We have analyzed the $\nu p$--process in core-collapse supernovae by taking into account the enhanced $3\alpha$ reaction rates induced by inelastic scatterings of free neutrons and protons on the Hoyle state. We find a negligible impact  from the neutron-induced inelastic scattering but a large effect from the proton-induced scattering. For the SN neutrino-driven wind, the particle-induced Hoyle state de-excitation suppresses the  production of $^{92,94}$Mo and $^{96,98}$Ru  abundances as reported in previous work.  On the other hand, for the HN wind with a massive PNS $\sim3M_{\odot}$,  $^{92,94}$Mo and $^{96,98}$Ru are enhanced by the Hoyle state effect although the nuclear yields for $A>140$  are reduced. 

The calculated abundance yields of the HN wind model were then applied to the GCE calculation of $p$--nuclei.  We found that the HN $\nu p$--process significantly contributes to the GCE of $^{92,94}$Mo and $^{96,98}$Ru, regardless of the particle-induced Hoyle state de-excitation. We have  demonstrated a possible contribution  from the HN $\nu p$--process to the origin of $p$--nuclei in the solar system.  Further studies, including an analysis of uncertainties neglected in our calculation, could  help substantiate these conclusions.

\section*{Acknowledgement}
This work was supported in part by Grants-in-Aid for Scientific Research of Japan Society for the Promotion of Science (19J13632, 20K03958, 21J11453). Work at the University of Notre Dame (GJM) was supported by DOE nuclear theory grant DE-FG02-95-ER40934. This work (TK) was also supported in part by the National Key R\&D Program of China (2022YFA1602401).


\bibliography{ref_sasaki,ref_YY}{}
\bibliographystyle{aasjournal}



\end{document}